\documentclass[aps,prl,twocolumn,letterpaper,reprint]{revtex4-1}

\usepackage{graphicx}     

\newcommand{\du}{\mbox{$\,$}\mathrm{d}}
\newcommand{\rvec}[2]{\mathbf{r}_{#1}^{#2}}
\newcommand{\qvec}{\mathbf{q}}
\newcommand{\rpar}{\mathbf{x}}

\newcommand{\gammainf}{\gamma_{\infty}}
\newcommand{\ellav}{\ell}
\newcommand{\Eq}[1]{Eq.~(\ref{eq:#1})}

\begin{document}

\addtolength{\topmargin}{0.125in}

\addtolength{\oddsidemargin}{0.00in}

\pagestyle{empty}


\onecolumngrid
\begin{center}

{\bf \large Capillary Fluctuations and Film--Height--Dependent
Surface Tension \\ of an Adsorbed Liquid Film}

\end{center}

\vspace{-0.25in}


\onecolumngrid
\begin{center}

Luis G. MacDowell{$^{1}$},
Jorge Benet{$^{1}$} and
Nebil A. Katcho{$^{2}$}
\end{center}


\vspace{-0.25in}

\onecolumngrid
\begin{center}

{\it $^1$Departamento de Qu\'imica F\'isica, Facultad de Qu\'imica,
  Universidad Complutense de Madrid, 28040  Madrid, Spain}

{\it $^2$ LITEN, CEA-Grenoble, 17 rue des Martyrs, 38054
Grenoble Cedex 9, France.}

\end{center}


\onecolumngrid
\vspace{0.0in}
\vskip 10pt

{\samepage
{\bf
\begin{center}
Abstract
\end{center}
}
\begin{center}
\begin{minipage}{.8\textwidth}
{\small 
Our understanding of both structure and dynamics of adsorbed liquids
heavily relies on the capillary wave Hamiltonian, but
a thorough test of this model is still lacking.  
Here we study the capillary wave fluctuations of a liquid film with short range 
forces adsorbed on a solid exhibiting van der Waals interactions.  We show 
for the first time that the measured capillary wave spectrum right above 
the first order wetting transition provides an interface potential 
consistent with independent calculations from thermodynamic integration. 
However, the surface tension exhibits an oscillatory film 
thick dependence which reveals a hitherto unnoticed
capillary wave broadening mechanism beyond mere interfacial displacements.
}
\end{minipage}
\end{center}
}

\twocolumngrid

State of the art nanotechnology allows for the synthesis of intricate
devices, with microchannels, grooves and minute containers that offer
a great number of potential applications.\cite{burns96} Taking advantage of
this synthetic technology also requires a thorough understanding and control
of the fluid behavior within the structured materials.\cite{seemann05} In
surface physics, most of our understanding of adsorbed fluids relies 
on the concept of {\em interface potential}, $g(\ell)$, the (surface)
free energy of a flat adsorbed liquid film of height $\ell$ above the
substrate.\cite{dietrich88,schick90} For practical applications,
however, we expect that adsorbed condensates within a device
will display a complex
geometry which can no longer be described by a simple scalar film height, but
rather, by a complex film profile $\ell({\bf x})$ which is now a
function of the position ${\bf x}$ on the plane  of the substrate.  
Our understanding of such
situations heavily relies on the phenomenological capillary wave 
Hamiltonian, $H[\ell]$ (CWH), a functional of the film profile which 
assumes a free
energy $g(\ell({\bf x}))$ locally, 
but includes extra
contributions due to  bending of the interface
via the liquid--vapor surface tension, $\gammainf$:\cite{buff65,vrij66}
\begin{equation}\label{eq:cwh}
  H[\ell] = \int \du \rpar \left (
   g(\ell(\rpar)) +  \gammainf \sqrt{1 + (\nabla \ell)^2}
   \right )
\end{equation} 
The equation above is at the heart of most theoretical accounts of surface 
phenomena, including, renormalization group analysis of wetting 
transitions \cite{fisher85}, the study of droplet profiles \cite{degennes04},
or the calculation of line tensions \cite{dobbs93}. 

The validity of the CWH, however, very much depends on dimensionality 
and range of the intermolecular forces considered. In two dimensions,
the CWH may be recovered exactly from the Ising model.\cite{upton02} 
For three dimensions and short--range forces, on the other hand, it has been 
long recognized that \Eq{cwh} cannot be derived bottom--up from a microscopic 
Landau--Ginzburg--Wilson functional (LGW) even for a flat substrate.\cite{jin93,parry06} 
Fisher and Jin argued that the form of \Eq{cwh} may
be retained provided one employs a modified local interface potential and 
a position dependent surface tension.\cite{jin93} For the short--range critical
wetting transition, however, Parry and collaborators have shown that \Eq{cwh}
has to  be discarded altogether in favor of a nonlocal functional of
$\ell(\rpar)$, which does recover the result of Fisher and Jin
in the case of complete wetting.\cite{parry06} Surprisingly, the very practically relevant
case of a flat substrate subject to long--range wall--fluid interactions
exhibiting a first order wetting transition has
received much less attention.\cite{bernardino09} It is this problem that we
would like to address in this Letter.

A simple means of testing \Eq{cwh} is to consider a film adsorbed on a
flat substrate.\cite{tidswell91,doerr99}  Thermal excitations of the otherwise planar film  of
lateral area $A$,
produce capillary waves, with amplitudes that are damped by the interface 
potential.\cite{vrij66} Expanding the film profile in Fourier modes,
$\ell(\rpar)=\sum_{\qvec} \ell_{\bf q} e^{i\qvec\rpar}$, allows us to
calculate a capillary wave spectrum (CWS).  
In the regime of Gaussian fluctuations, the CWS expected from 
\Eq{cwh} is:\cite{rowlinson82b,mecke99b,fernandez12}
\begin{equation}\label{eq:cws}
   \frac{k_BT}{A<|\ell_{\bf q}|^2>} = g''_{\rm cws} + \gamma_{\rm
   cws}\, q^2 + 
   \kappa_{\rm cws}\, q^4
\end{equation} 
where $k_B$ is the Boltzmann constant, $T$ is the temperature and
the angle brackets denote a thermal average; while
$g''_{\rm cws}$, $\gamma_{\rm cws}$ and $\kappa_{\rm cws}$ are the
first few coefficients of an expansion in powers of ${\bf q}$.
 Even for Gaussian fluctuations, the relation between the coefficients of \Eq{cwh} and
\Eq{cws} may not be trivial. In fact, \Eq{cws} provides  at 
best the renormalized interface potential and surface tensions.  In our three--dimensional 
system subject to long--range interactions, however,  we are safely above the
upper critical dimension, and do not expect significant
renormalization.
The first two terms in the right--hand side, with $g''_{\rm
cws}=g''(\ell)$ and $\gamma_{\rm cws}=\gammainf$ then follow  from the
classical model of \Eq{cwh}, and may be measured independently, 
while an extra term of order $q^4$ is known
to arise already for free liquid--vapor interfaces in the absence of
an external field.\cite{mecke99b,fernandez12} 

Here we show for the first time that $g''_{\rm cws}$ as obtained from the CWS, 
is fully consistent with independent results of $g(\ell)$  calculated
via
thermodynamic integration; while an extra film--height dependence of the effective stiffness,
$\gamma_{\rm cws}=\gammainf+\Delta\gamma(\ell)$, 
with $\Delta\gamma(\ell) \propto g''(\ell)$ holds for films
up to a few molecular diameters thick.

\begin{figure}[t]
\centering
\includegraphics[clip,scale=0.30,trim=0mm 5mm 0mm 0mm,
angle=270]{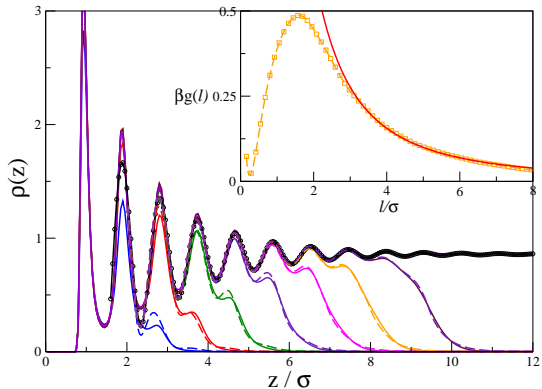}
\caption{Density profiles for adsorbed layers of thickness $\ellav$
of (full lines, from left to right) 1.9, 3.0, 4.1, 5.3, 6.5, 7.3 and 8.5 
$\sigma$. The symbols depict the function $A_w \cos(k_o z + \theta) \exp(-b_o z)$
with $k_o$ and $b_l$ obtained from the bulk correlation function, and
$A_w$, $\theta$ obtained from a fit to an adsorbed layer of thickness
$\ellav$ ca. 35 $\sigma$. Dashed lines are predictions from the
superposition model, \Eq{idp}. 
Inset: Interface potential as a function of average film thickness, 
$\ellav$ (symbols).  Expectations from the Hamaker model are shown as
a dark line.}
\label{profiles}
\end{figure}

Our simulations are performed for a well--known model of ``Argon
adsorbed on solid carbon dioxide" that was employed extensively
in order to confirm the  theoretical prediction of prewetting by Ebner 
and Saam.\cite{finn88,sokolowski90,ebner77}  
Argon is described by means of a Lennard--Jones potential, with
a well depth of $\epsilon$ and a molecular diameter of $\sigma$ 
(henceforth employed as length unit). Note, however,
that the fluid--fluid interactions are short range,
since the Lennard--Jones potential is truncated beyond a distance of
$2.5\sigma$.  The fluid interacts above the upper half plane $z>0$ with a 
flat inert substrate  exhibiting a truly long range potential
of the form $V(z)=-H_w/z^3$, with $H_w$ the Hamaker constant and $z$ the
perpendicular distance to the substrate.\cite{supplemental13}
This choice of interactions turns out
to be of great advantage. On the one hand, 
we can exploit various mean field results that are known for
systems with short--range fluid--fluid
forces.\cite{chernov88,evans94,henderson94,jin93,parry06} 
On the other hand,
the long range external field amplifies the signature of capillary waves 
that would otherwise pass unnoticed. This
will allow us to test predictions of
relevance to short--range--wetting that have remained a matter of
debate for many years.\cite{jin93,parry06}

We perform our study at a temperature $k_BT/\epsilon=0.60$, just above
the wetting transition 
(ca. $k_BT_w/\epsilon=0.598$).\cite{errington04,gregorio12}
Close to coexistence, an adsorbed film of  liquid
density, $\rho_l$, is found in equilibrium with a vapor phase,
$\rho_v$.
Direct measurements of the interface potential are obtained by
performing grand canonical simulations 
and collecting the
probability, $P(\Gamma)$ of finding a given adsorbed amount $\Gamma$ of 
molecules.\cite{supplemental13}
Using the intrinsic sampling method
(ISM),\cite{tarazona04}
an accurate technique for locating the interface of a corrugated film,
we relate adsorption to average film thickness, $\ellav$.
A coarse--grained interface potential is then estimated as 
$g(\ellav) = -k_BT\ln P(\ellav)$.\cite{macdowell06,macdowell11,gregorio12}

In our system, the wetting temperature is very low, close to the fluid's 
triple point.  Hence,
the density profile of a an  adsorbed liquid film is dominated by a single
relaxation mode  with damped oscillatory
behavior,\cite{chernov88}
such that $\rho(z)=\rho_l [1 + h(z)]$, with $h(z)=A_w \cos(k_o z + \theta)
e^{-b_o z}$. We checked that both the inverse correlation length
$b_o$ and the wave vector $k_o$ are equal to those required
to describe the oscillations of the bulk liquid
correlation function (c.f. Fig.\ref{profiles}), as expected for
short--range fluids subject to
an external field. Only the amplitude, $A_w$ and phase $\theta$ of the
oscillations are dictated by the external 
field.\cite{evans94,henderson94,klapp08b}
The mean field interface potential emerging from this scenario
may be split into intrinsic and external field contributions, 
$g(\ellav) = g_{mb}(\ellav) + g_V(\ellav)$. 
The first term, $g_{mb}(\ellav)$ is a highly non--trivial functional of
the number density $\rho(z;\ellav)$, and includes many--body fluid--fluid
correlations. Whereas its structure may be quite complex,
and could in principle pick up oscillatory behavior of the
adsorbed layer  it is expected to decay exponentially
fast.\cite{chernov88} The second contribution,
\begin{equation}\label{eq:g_ext}
  g_V(\ellav) = \int V(z) \rho(z;\ellav) \du z
\end{equation} 
stems from the wall--fluid external field
and recovers the familiar long--range Hamaker dependence 
$g_V(\ellav) \propto \ellav^{-2}$.

\begin{figure}[t]
\centering
\includegraphics[clip,scale=0.30,trim=0mm 5mm 0mm 0mm,angle=270]{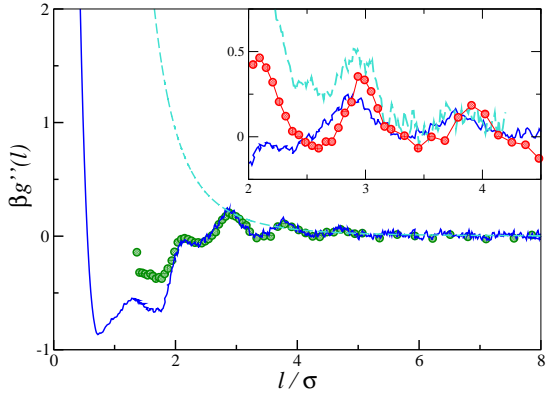}
\caption{Second derivative of the interface potential (full line)
compared with expectations from the CWS, $g_{cws}''$ (green circles)
and from the Hamaker model (dashed line). Notice almost perfect match
of $g''$ with $g_{cws}''$ up to $\ellav=2\sigma$. Inset: Enlarged view
of $g''$ (full line), as compared with $g''_V$ (dashed line) and
$\gamma_{cws}-\gamma_{\infty}$ (symbols).
\label{d2g2}}
\end{figure}

The coarse grained interface potential that is obtained in our
simulations is renormalized up to wavelengths equal to the lateral
system dimensions, and shows no sign of oscillatory behavior
(Fig.\ref{profiles}).  Rather, beyond about $\ellav=3\sigma$
it exhibits a monotonic decay
that is usually attributed to long--range forces.
The signature of layering is revealed only when we consider the
derivatives of $g(\ellav)$. Particularly, 
$g''(\ellav)$, which we can evaluate numerically thanks to our highly
accurate data (Fig.\ref{d2g2}) shows a clear oscillatory behavior 
superimposed
on the expected long--range decay.  
In order to 
unravel the nature of these oscillations, we evaluated \Eq{g_ext}
numerically, using simulated density profiles for $\rho_{\ellav}(z)$. The
results are indistinguishable from the long--range Hamaker contribution
$\approx \frac{1}{2} H_w \Delta \rho\, \ellav^{-2}$
depicted as a dashed line in 
Fig.\ref{profiles}. However, taking second derivatives, we
observe once again a clear oscillatory behavior of 
$g_V''(\ellav)$, that
resembles the oscillations of $g''(\ellav)$ appearing between about
two and four $\sigma$ (Fig.\ref{d2g2}). This shows that not only
the decay of $g''$, but also the oscillations are to a great
extent due to the external field contribution.

\begin{figure}[t]
\centering
\includegraphics[clip,scale=0.30, trim=0mm 5mm 0mm 0mm,
angle=270]{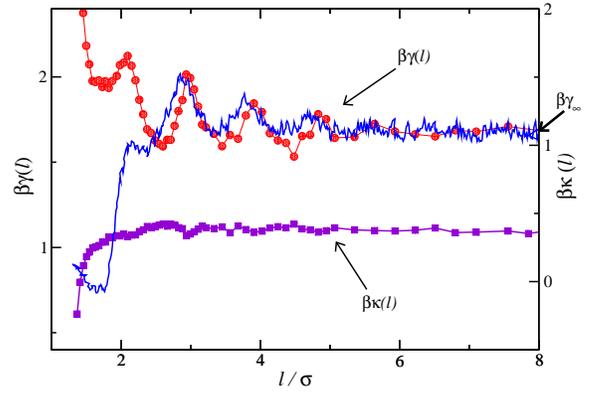}
\caption{Film--thick--dependent surface tension, $\gamma_{cws}(\ellav)$
(circles, left axis) and bending rigidity $\kappa_{cws}$ 
(squares, right axis) as
obtained from the CWS. The full lines are predictions from the model
$\gamma_{\infty}+g''(\ell)/b^2$, with $b=0.85~\sigma^{-1}$.
The thick arrow  points to the value of
$\gammainf$ as obtained independently for the free interface.
\label{gamma}
}
\end{figure}

Independent canonical simulations are carried out for adsorbed films 
ranging from one to about ten molecular diameters thick. For each
configuration a film--height profile $\ell(\rpar)$ is 
determined,\cite{tarazona04} and the Fourier components $\ell_{\qvec}$ 
calculated. The thermal average $<|\ell_{\qvec}|^2>$ is fitted 
to a
quadratic polynomial in $q^2$, as suggested by expectations from
\Eq{cws}, yielding independent estimates of $g''_{\rm cws}$,
$\gamma_{\rm cws}$ and $\kappa_{\rm cws}$ (Fig.\ref{d2g2}--\ref{gamma}).

The results for $g''_{\rm cws}$ as obtained from the CWS, show an
excellent agreement with independent estimates of $g''$ 
obtained from thermodynamic
integration, for all film lengths up to 2~$\sigma$ (Fig.\ref{d2g2}).
More impressively, the good agreement holds even in regions where
$g''<0$. Previously,
interface potentials in the unstable region have been estimated by
studying the behavior of dewetting patterns.\cite{kim99,seemann01b,herring10}
Films with negative  $g''$ are unstable to all
perturbations beyond the critical wavelength
$\lambda_c = 2\pi \left ( \gamma_{\infty}/|g''| \right
)^{1/2}$.\cite{vrij66,seemann01b}
Using  $\beta\gammainf=1.66\sigma^{-2}$,\cite{gregorio12}
and the value of $g''$ at the second minima, where the agreement
breaks down, we find that the onset of
instability occurs for fluctuations of wavelength $\lambda_c\approx 10\sigma$.
This  is just about the lateral system size of our study, 
$L=10\sigma$ and serves to motivate why we are able to equilibrate ``unstable" 
films in our finite size simulations.  
Unfortunately, the cluster criteria required to define the Fourier 
modes $\ell_{\qvec}$ breaks down below $\ellav=\sigma$, so that we are unable 
to compare $g''$ and $g_{cws}''$  in this range.

According to \Eq{cws}, the first order 
coefficient in our fit to $1/<|\ell_{q}|^2>$ yields an estimate of the 
stiffness. This expectation has been confirmed 
for a free liquid--vapor interface in several 
studies.\cite{mueller96b,lacasse98,milchev02,tarazona04,vink05} 
In our adsorbed films we find  good agreement with
independent measures of  $\gammainf$.\cite{gregorio12} For films below about 
$\ellav<6\sigma$, 
however, $\gamma_{cws}$  becomes film--thick 
dependent and picks up a strong oscillatory behavior vaguely resembling that 
of $g''$ (Fig.\ref{gamma}).  
 Our results show that,
as opposed to $\gamma_{cws}$,
the bending rigidity, $\kappa_{cws}$ is rather insensitive to $\ellav$, 
remaining almost constant up
to $2\sigma$, where it strongly decreases and becomes negative
(Fig.\ref{gamma}) as the film becomes unstable (i.e., $g''(\ellav)<0)$).

Here we would like to point out that the need for a film--thick
dependent stiffness results naturally from an improved
theory of capillary wave broadening incorporating 
distortions of the intrinsic density profile.

First we consider the liquid--vapor interface for an adsorbed
film of infinite thickness, obeying a LGW density
functional.
We seek density profiles minimizing the free energy functional 
subject to the constraint,  
$\rho_{lv}(\rpar,z=\ell(\rpar);\Sigma)=\rho_{1/2}$, 
where $\rho_{1/2}=\frac{1}{2}(\rho_l + \rho_v)$ (crossing criterion)
and $\Sigma$ is shorthand for the functional dependence,
$[\ell(\rpar)]$. A general solution to this problem is not possible,
but  treating liquid and vapor branches of the density profile
separately (double parabola approximation) leads to the
following Helmholtz equation as the solution to the functional
minimization:\cite{jin93,parry06}
\begin{equation}
   \nabla^2 \Delta \rho_{lv}(\rvec{}{};\Sigma) - b^2 \Delta
   \rho_{lv}(\rvec{}{};\Sigma) = 0
\end{equation} 
where
$\Delta\rho_{lv}(\rvec{}{};\Sigma)=\rho_{lv}(\rvec{}{};\Sigma)-\rho_{\infty}$ is the
density excess over the bulk value, $\rho_{\infty}$, we assume
liquid (vapor) asymptotic densities to the left (right) of $z=\ell$,
and $b$ is the decay rate of the
liquid--vapor interface. Considering an expansion of the density
profile in transverse Fourier modes, $\Delta\rho_{lv}(z;\qvec{}{})
e^{i\qvec{}{}\cdot \rpar{}{}}$,
 we find a
general solution of the form:
\begin{equation}\label{eq:general}
 \rho_{lv}({\bf r};\Sigma) = \rho_{\infty} + \sum_{\bf q} A_{\bf q} e^{\pm b_q z}
 e^{i{\bf q}\cdot {\bf x}}
\end{equation} 
where $\pm$ indicates solutions for either the left (liquid) or right
(vapor) half planes; while
the eigenvalues $b_q=\sqrt{b^2 + q^2}$ physically correspond
to wave vector dependent inverse correlation lengths, that impose fast
damping of the density perturbation for short wavelengths. 

Now consider the liquid--vapor interface as it approaches the substrate.
A smectic density wave of the form $\rho_l (1 + h(z))$ propagates from the
wall outwards and perturbs the asymptotic liquid density, $\rho_l$.
For thin films, the  liquid--vapor interface feels
the distortion, and is thus modulated by $h(z)$.  Accordingly,
we consider a {\em superposition model}, whereby the density
profile  of the adsorbed film results from the superposition of
$h(z)$ onto $\rho_{lv}(\rvec{}{};\Sigma)$, i.e.,
$\rho(\rvec{}{};\Sigma) = [1 + h(z) ] \rho_{lv}(\rvec{}{};\Sigma)$. 
In order to obtain the Fourier coefficients, $A_{\bf q}$ of the
distorted liquid--vapor interface, we evaluate $\rho({\bf r},\Sigma)$ 
at the boundary, $z=\ell({\bf x})$.
Assuming a flat interface, with
$\ell(\rpar{}{})=\ellav$ everywhere, we readily obtain
the zero order result for $A_{{\bf q}={\bf 0}}$, which yields
the following {\em intrinsic} density profile:
\begin{equation}\label{eq:idp}
 \rho_{\pi}(z;\ellav) = \left [ 1 + h(z) \right  ] \left ( \rho_{\infty}  +
  \frac{\Delta\rho_{1/2} - \rho_{\infty} h(\ellav)}{1 +
  h(\ellav)} e^{\pm b(z-\ell)} \right )
\end{equation} 
This simple model already provides a rather good description of the
average density profiles, $\rho(z;\ell)$ obtained from simulation (c.f
Fig.\ref{profiles}).

In order to assess the role of capillary waves
we follow 
a procedure suggested by Tarazona,\footnote{P. Tarazona, 
private communication.} first performing a Taylor expansion about 
deviations from a planar interface 
$\delta \ell({\bf x}) = \ell({\bf x}) - \ell$ and  then Fourier 
transforming  the resulting expression. Retaining terms up
to order $|\delta \ell_{\qvec{}{}}|^2$, and integrating over
${\rpar}$, we find that
the laterally averaged density profile,
$\rho(z;\Sigma)=<\rho({\bf r};\Sigma)>_{\bf x}$
may be written
solely in terms of the intrinsic density profile as:
\begin{equation}\label{eq:ncwb}
\rho(z;\Sigma)   = 
\rho_{\pi}(z;\ell) + \frac{1}{2} \sum_{\bf q} \left [
   \frac{d^2 \rho_{\pi}(z)}{d \ell^2} \mp
    \frac{d \rho_{\pi}(z)}{d \ell} \frac{q^2}{b}
 \right ] | \delta \ell_{\bf q}|^2   
\end{equation} 
This result features capillary wave broadening effects
beyond the classical theory, which includes only the first two terms
in the right hand side. Here we see that broadening of the profile
is enhanced by terms $<(\frac{\nabla \ell}{b})^2>_x$ 
that account for distortion of the profile due to curvature 
of the interface.

In order to understand the significance of this contribution most easily,
it is convenient to consider the limit of thick films, where 
 $d^2 \rho_{\pi}(z) /d \ell^2$ and $\mp b d \rho_{\pi}(z)/d \ell$ 
become identical in the double parabola approximation. Substitution of
\Eq{ncwb} into \Eq{g_ext}, immediately yields the following external field
contribution to the interface potential,
\begin{equation}\label{eq:ncwt}
g_{V}(\Sigma) = g_V(\ellav) + \frac{1}{2} g_V''(\ellav) \sum_{\bf q} ( 1 +
\frac{q^2}{b^2} ) |\delta \ell_{\bf q}|^2
\end{equation} 
thus, the enhanced broadening of the density profile results in an
extra contribution of order $q^2$ that is absent in the classical
theory.

Employing this result for the interface potential in \Eq{cwh}, 
we find that the CWS must now take the form:
\begin{equation}\label{eq:ncws}
  \frac{k_BT}{A<|\ell_{\qvec}|^2>} = 
   g''(\ell) + \left ( \gammainf + \frac{g''_V(\ell)}{b^2} \right ) q^2
\end{equation} 
i.e., the distortion of the density profile due to capillary waves
couples $g_V(\ellav)$ with terms of order $q^2$, effectively
providing a film--thick--dependent surface tension. According
to our analysis, the $\ellav$ dependence will closely follow
$g''(\ellav)$ as long as the film is thick enough and the interface potential 
is dominated by the external field contribution.  Particularizing
\Eq{ncwt} and \Eq{ncws} for wall--fluid interactions with algebraic
decay our results are exactly as obtained previously 
from the nonlocal theory of interfaces,\cite{bernardino09} 
and
thus lend support to to the nonlocal theory of short--range--wetting.\cite{parry06}

We do not expect this simplified treatment to provide 
a quantitative description of our simulations since 
i) the fluid actually has different correlation lengths for the
liquid and vapor phases
and ii) the superposition model does not accurately
describe the density derivatives of the film
profile with respect to $\ell$.
However, we can test the predictions for 
$\gamma_{\rm cws}\approx\gammainf + g''(\ell)/b^2$, 
using our independent results for 
$\gammainf$ and $g''(\ellav)$, and considering $b$ as an 
effective inverse correlation length.
Surprisingly, a test of this hypothesis provides good agreement for films of
thickness down to $\ellav=2.5\sigma$, where $h(z)$ is still far from
having decayed  (Fig.\ref{gamma}).  Interestingly,  the steep increase of
$\gamma_{\rm cws}(\ellav)$ below $\ell_0=2.5\sigma$, which is not captured 
by the 
full $g''(\ellav)$, rather obeys qualitatively the trend followed 
by  $g_V''(\ellav)$ as obtained from integration of \Eq{g_ext} 
(Fig.\ref{profiles}).

The significance of the effective surface tension may be readily
understood in terms of the parallel correlation length, which is
defined in the classical theory as 
$\xi_{\parallel}^2=\gamma_{\infty}/g''(\ell)$. Using instead the
effective surface tension we find that
$\xi_{\parallel}^2=\gamma_{\infty}/g''(\ell)+\xi_b^2$, where $\xi_b$ is the
bulk correlation length. As the classical result may become small
either close to a critical point or under a strong field, our
improvement 
merely provides the fairly obvious requirement of a lower bound 
$\xi_b$ to $\xi_{\parallel}$. 
This expectation may be 
assessed experimentally for a free fluid--fluid interface under the
gravitational field, since  then 
$\xi_{\parallel}^2=a^2+\xi_b^2$, with $a$ the capillary length.
A colloid/polymer suspension  close to the fluid--fluid consolute point
seems an excellent candidate for experimental verification.\cite{aarts04}

In summary, we have demonstrated that capillary wave fluctuations of
thick adsorbed liquid films obey expectations from classical capillary 
wave theory. For thin films, however, a hitherto unnoticed capillary 
wave broadening mechanism results in an effective film--thick--dependent surface
tension. This observation has important implications
in both the structure and dynamics of adsorbed condensates.

LGM would like to thank P. Tarazona for 
helpful discussions and  critical reading of the manuscript as well as
E. M. Fern\'andez and E. Chac\'on for invaluable assistance
with the ISM. We also wish to thank  R. Evans, A. Parry, S. Dietrich, 
N. Bernardino,  V. Martin--Mayor and A. Rodriguez--Bernal 
for helpful discussions.  We acknowledge financial support from grant 
FIS2010-22047-C05-05 of the 
Spanish Ministerio de Economica y Competitividad, and project P2009/ESP/1691 
(MODELICO) from Comunidad Autonoma de Madrid.


%

\end{document}